\documentstyle[10pt]{article}

\newcommand\beq{\begin{equation}}
\newcommand\eeq{\end{equation}}
\newcommand{\eqref}[1]{(\ref{#1})}

\def\beqa{\begin{eqnarray}}
\def\eeqa{\end{eqnarray}}
\def\ba{\begin{array}}
\def\ea{\end{array}}
\def\r{\rangle}
\def\l{\langle}

\def\b{\beta}

\def\de{\mbox{det}}

\def\eps{\epsilon}

\def\sl{\sum\limits}

\def\lt({\left(}
\def\rt){\right)}
\def\pdx{\frac{\partial}{\partial x}}

\def\intpi{\int\limits_{-\pi}^{\pi}}
\def\intla{\int\limits_{-\Lambda}^{\Lambda}}
\newcommand{\bl}[1]{\makebox[#1em]{}}

\begin{document}
\begin{titlepage}
\title{\Huge{Correlators of the phase model}}
\author{N.M. Bogoliubov
\thanks{Supported by the Academy of Finland}\ \thanks{On leave of absence from St. Petersburg Department of the Steklov Mathematical Institute,
Fontanka 27, St. Petersburg 191011, RUSSIA,
 e-mail: bogoliub@tftxb.helsinki.fi}
\\
Research Institute for Theoretical Physics\\
P.O.Box 9, FIN-00014 University of Helsinki, FINLAND.\\ \\
A.G. Izergin\thanks{Supported in part by MAE-MICECO-CNRS Fellowship}
\thanks{On leave of absence from St. Petersburg Department of the Steklov Mathematical Institute,
Fontanka 27, St. Petersburg 191011, RUSSIA, e-mail: aizergin@enslapp.ens-lyon.fr}
,\bl{1} N.A. Kitanine \thanks{Supported in part by project MAE 96/9804}\ 
\thanks
{On leave of absence from St. Petersburg Department of the Steklov Mathematical Institute,
Fontanka 27, St. Petersburg 191011, RUSSIA, e-mail: Nikolai.Kitanine@enslapp.ens-lyon.fr}
\\
Laboratoire de Physique Th\'eorique
{\sc enslapp}\thanks{URA 14-36 du CNRS,
associ\'ee \`a l'E.N.S. de Lyon,
et \`a l'Universit\`e de Savoie}
\\
ENSLyon,
46 All\'ee d'Italie,\\
69007 Lyon, FRANCE\\
\\} 

\date{October 1996}
\maketitle

\begin{abstract}
 We introduce the phase model on a lattice and solve it 
using the algebraic Bethe ansatz. Time-dependent temperature
correlation functions of phase operators and the "darkness
formation probability" are calculated in the thermodynamical limit. 
These results can be used to construct integrable equations for
the correlation functions and to calculate their asymptotics.
\end{abstract}
\vskip 1cm
\rightline{{\small E}N{\large S}{\large L}{\large A}P{\small P}-L-622/96}
\rightline{HU-TFT-96-41}
\end{titlepage}

 Phase operators were intensively studied in quantum optics
\cite{cn1,cn2,ly}. They can be defined by the following commutation
relations
\beq
\label{comm}
\lbrack N,\phi^{+}]=\phi^{+},
\makebox[1em]{}[N,\phi]=-\phi,\bl{1} \lbrack
\phi,\phi^{+}]=\pi,
\eeq
where $\pi$ is the vacuum projector
\[\pi=(|0\r\l0|)\]
and $N$ is the number of particles operator.

The phase model is a model of interacting phase operators
on a lattice. It was constructed in \cite{bbpt} as a limit 
case of the 
$q$-boson hopping model \cite{bb,bbp} . The Hamiltonian of the
phase model has the following form
\beq
\label{ham}
H=-\frac 12\sum_{n=1}^M(\phi_n^{+}\phi_{n+1}+\phi_n\phi_{n+1}^{+}-2N_n),
\eeq
 where operators $\phi_j,\phi_j^+,N_j$ commute in the different sites
and satisfy the commutation relations \eqref{comm} in the same site.

 The complete set of eigenvectors for the model can be obtained
by means of the algebraic Bethe ansatz (see \cite{bik}
and references therein). $L$-operator
of the model has the form
\begin{equation}
\label{loph}
L_n(p)=\left(
\begin{array}{cc}
e^{ip/2} & \phi _n^{\dagger } \\
\phi _n & e^{-ip/2}
\end{array}
\right) ,
\end{equation}
This operator
satisfies the bilinear relation
$$
R(p,s)L_n(p)\otimes L_n(s)=L_n(s)\otimes L_n(p)R(p,s),
$$
in which $R(p,q)$ is the 4$\times $4 matrix $R$-matrix. The non-zero
elements of this $R$-matrix are%
$$
R_{11}(p,s)=R_{44}(p,s)=f(s,p),
$$
$$
R_{22}(p,s)=R_{33}(p,s)=g(s,p),
$$
$$
R_{23}(p,s)=1
$$
and
\begin{equation}
\label{fgph}f(p,s)=i\frac{e^{i\frac{p-s}2}}{2\sin (\frac{s-p}2)}%
;\,\,g(p,s)=\frac i{2\sin (\frac{s-p}2)}.
\end{equation}
The Bethe equations for the model
\begin{equation}
\label{betheph}
\exp \{i(M+N)p_j\}=(-1)^{N-1}\exp \{i\sum_{k=1}^Np_k\}
\end{equation}
$(j=1,...,N)$ are exactly solvable:
\begin{equation}
\label{solph}
p_j=\frac{2\pi I_j+\sum_{k=1}^Np_k}{M+N}\,,
\end{equation}
where $I_j$ are integers or half-integers depending on $N$ \thinspace being
\thinspace odd or even.

The $N$-particle eigenenergies of the Hamiltonian $H_\mu 
=H-\bar \mu \hat N$ (\ref
{ham}) are
\begin{equation}
\label{enph}
E_N=\sum_{k=1}^N(h(p_k)-\bar \mu );\,\,h(p)=2\sin {}^2(p/2).
\end{equation}
Here $\bar \mu $ is the chemical potential, $0\leq \bar \mu \leq 1.$

The thermodynamics of the model we shall consider for the case when the
total momentum $P=\sum_{k=1}^Np_k$ is zero $P=0.$

The thermodynamics of the model is handled in the standard way. The ground
state energy of the model at finite temperatures $\beta ^{-1}$ is determined
through the solution of the nonlinear integral equations
\begin{equation}
\label{eenph}\epsilon (p)=h(p)-\bar \mu -(2\pi \beta )^{-1}\int\limits_{-\pi }^\pi
\ln (1+e^{\beta \epsilon (p)})dp,
\end{equation}
$$
2\pi \rho (p)(1+e^{\beta \epsilon (p)})=1+\int\limits_{-\pi }^\pi \rho (p)dp.
$$
The function $\rho (p)$ is a quasi-particle density while $\epsilon (p)$ is
the excitation energy. The pressure is then
\begin{equation}
\label{presph}{\cal P}=(2\pi \beta )^{-1}
\int\limits_{-\pi }^\pi \ln (1+e^{\beta
\epsilon (p)})dp
\end{equation}
and the density is
\begin{equation}
\label{denph}D=\frac{\partial {\cal P}}{\partial \bar \mu }=
\int\limits_{-\pi }^\pi
\rho (p)dp.
\end{equation}
So we have
\begin{equation}
\label{eneph}\epsilon (p)=h(p)-\bar \mu -{\cal P}
\end{equation}
and the quasi-particle density has the Fermi-like distribution
\begin{equation}
\label{disph}2\pi \rho (p)=(1+D)(1+e^{\beta \epsilon (p)})^{-1}.
\end{equation}

At zero temperature $\beta ^{-1}=0$ the ground state is the Fermi sphere $%
-\Lambda \leq p\leq \Lambda $ ($\Lambda \leq \pi )$ filled by the particles
with the negative energies $\epsilon _0(p).$ The pressure and density are now%
$$
{\cal P}_0=-(2\pi )^{-1}\int\limits_{-\Lambda }^\Lambda \epsilon
_0(p)dp,\,\,D_0=\int\limits_{-\Lambda }^\Lambda \rho_0 (p)dp.
$$
\thinspace From (\ref{eneph}) and (\ref{disph}) we have
$$
\epsilon _0(p)=h(p)-\bar \mu -{\cal P}_0,\,\,\epsilon _0(\pm \Lambda )=0;
$$
$$
2\pi \rho _0(p)=(1+D_0).
$$
From the last equation we can express the bare Fermi momentum $\Lambda $ as
the function of density%
$$
\Lambda =\frac{\pi D_0}{1+D_0}.
$$
The Fermi velocity $v$ is equal:%
$$
v=\frac{\epsilon _0^{\prime }(\Lambda )}{2\pi \rho _0(\Lambda )}%
=(1+D_0)^{-1}\sin \frac{\pi D_0}{1+D_0}.
$$

When $\Lambda \rightarrow 0\,(\bar \mu \rightarrow 0),\,\,D_0\rightarrow 0$
and ${\cal P}_0\rightarrow 0$ as must be expected. When $\Lambda \rightarrow
\pi \,(\bar \mu \rightarrow 1)$ all the vacancies are now occupied by
particles $D_0\rightarrow \infty ,\,{\cal P}_0\rightarrow 1$ and the model (%
\ref{ham}) is the classical $XY$ chain in this limit \cite{bbp},\cite{bbt}.

 The correlation functions for the phase model can be represented as
Fredholm determinants of integral operators. The similar representations
were obtained recently for the model of impenetrable bosons \cite{l,iik}
and for the XX0 Heisenberg chain \cite{cikt}. It was shown that such results
can be used for calculation of the asymptotics of correlation functions.

 The simplest correlation function is the emptiness formation probability.
It can be defined as a probability of the states such that 
there are no particles
in the first $m$ sites of the lattice (see e.g. \cite{bik}). We will call it darkness formation probability in the case of the phase model.
Using the algebraic Bethe ansatz and the representation
for the scalar products of the Bethe states  \cite{k}
one can represent this function
as a Fredholm determinant
\beq
\label{efpT}
\tau(m,\b)=(1+D)\de(\hat{I}-\hat{M}),
\eeq
where $\hat{I}$ is the identity operator and
$\hat{M}$ is an integral operator 
\[(\hat{M}f)(p)= \int\limits_{-\pi}^{\pi}M(p,q)f(q)dq,\]
with the kernel
\beq
\label{kernT}
M(p,q)=\frac 1{2\pi}\sqrt{\nu(p,\b)}\frac{\sin\frac{m+1}{2}(p-q)}{\sin\frac{1}{2}(p-q)}
\sqrt{\nu(q,\b)},
\eeq
where $\nu(p,\b)=\lt(1+\exp(\b\eps(p))\rt)^{-1}$.

It can be easily shown using the representation \eqref{efpT}
that $\tau(m,\b)\leq 1$. 

At zero temperature one has 
\beq
\label{efpT0}
\tau_0(m)=(1+D_0)\de(\hat{I}-\hat{M}_0),
\eeq
where $\hat{M}_0$ is an integral operator
\beq
(\hat{M}_0f)(p)=\intla M_0(p,q)f(q)dq,
\eeq
with the kernel
\beq
\label{kernT0}
M_0(p,q)=\frac 1{2\pi}\frac{\sin\frac{m+1}{2}(p-q)}{\sin\frac{1}{2}(p-q)}.
\eeq

 The time-dependent correlation function of phase operators
can be also expressed as a Fredholm determinant. This representation
can be obtained using the representation for the form-factor
\cite{k}. We will consider
the temperature mean values
\beq
\label{corr+}
f^+(\b,m,t)=\l\phi_{m+1}(t)\phi_{1}^+(0)\r_\b,
\eeq
\beq
\label{corr-}
f^-(\b,m,t)=\l\phi_{m+1}^+(t)\phi_{1}(0)\r_\b,
\eeq
where 
\[\phi_{m}(t)=\exp[iH_\mu t]\phi_{m}\exp[-iH_\mu t].\]

This correlation function can be represented in the following form
\beq
\label{cortherm}
f^{(\pm)}(m,t,\b)=\exp\lt(\frac D{1+D}\rt)\frac{1+D}{2\pi}\sl_{l=0}^{\infty}
\intpi e^{i(m-l)\Theta}h^{(\pm)}(l,t,\b,\Theta)d\Theta.
\eeq
The functions $h^{(\pm)}(l,t,\b,\Theta)$ can be written as a Fredholm
determinant
\beq
\label{htherm}
h^{(\pm)}(l,t,\b,\Theta)=\lt( G(l,t)+\pdx\rt)\de(\hat{I}+
\hat{V}\mp x\hat{R^\pm})|_{x=0},
\eeq
where $\hat{V}$ and $\hat{R}$ are integral operators 
\beq
\ba{c}
(\hat{V}f)(p)=\frac 1{2\pi}\intpi V(p,q)f(q)dq,\\
(\hat{R^\pm}f)(p)=\frac 1{2\pi}\intpi R^\pm(p,q)f(q)dq,
\ea
\eeq
with kernels
\[V(p,q)=\frac {e^{-i\frac{p-q}{2}}}{\sin\frac 12 (p-q)}
(\frac 12 (E_+^+(l,t,p,\b,\Theta)+E_+^-(l,t,p,\b,\Theta))
E_-(l,t,q,\b)-\]
\beq
\label{VT}
-\frac 12 (E_+^+(l,t,q,\b,\Theta)+E_+^-(l,t,q,\b,\Theta))
E_-(l,t,p,\b)),
\eeq
\beq
\label{RT+}
R^+(p,q)=E_+^+(l,t,p,\b,\Theta)E_+^-(l,t,q,\b,\Theta),
\eeq
\beq
\label{RT-}
R^-(p,q)=E_-(l,t,p,\b)E_-(l,t,q,\b),
\eeq
where the functions $G(l,t),E_-(l,t,p,\b),E_+^+(l,t,p,\b,\Theta)$ and
$E_+^-(l,t,p,\b,\Theta)$ are defined as
\beq
\label{GT}
G(l,t)=\frac 1{2\pi}\intpi \exp(ilq-it\epsilon(q))dq,
\eeq
\beq
\label{ET}
E(l,t,p,\Theta)=\frac 1 {2\pi}\mbox{v.p.}\intpi\frac{\exp(ilq-it\epsilon(q))}
{\tan \frac 12 (q-p)}dq+\tan\frac 12 (p-\Theta)\exp(ilp-it\epsilon(p)) ,
\eeq 
\beq
\label{E-T}
E_-(l,t,p,\b)=\sqrt{\nu(p,\b)}\exp\lt(-i\frac l2 p+i\frac t2 \epsilon(p)\rt),
\eeq
\beq
\label{E++T}
\ba{c}
E_+^+(l,t,p,\b,\Theta)=\frac 12 E_-(l,t,p,\b)\times\\
\times \lt((E(l,t,p,\Theta)+e^{-i\Theta}E(l+1,t,p,\Theta))+i
(G(l,t)+e^{-i\Theta}G(l+1,t))\rt),
\ea
\eeq
\beq
\label{E+-T}
\ba{c}
E_+^-(l,t,p,\b,\Theta)=\frac 12 E_-(l,t,p,\b)\times\\
\times \lt((E(l,t,p,\Theta)+e^{i\Theta}E(l-1,t,p,\Theta))-i
(G(l,t)+e^{i\Theta}G(l-1,t))\rt).
\ea
\eeq

  Although the kernel \eqref{VT} seems to be very complicated
it has the form which allows to use the method proposed in
\cite{iik} for calculation of the asymptotics. 

 In the case $t=0, m>1$ the correlators
have the following form
\beq
f^{(\pm)}(m,0,\b)=\exp\lt(\frac D{1+D}\rt)\frac{1+D}{2\pi}
\sl_{l=m}^{\infty}\intpi e^{i(m-l)\Theta}h^{(\pm)}(l,0,\b,\Theta)d\Theta,
\eeq
\beq
\label{ht=0}
h^{(\pm)}(l,0,\b,\Theta)=\pdx\de (\hat{I}-\hat{v}+x\hat{r}^\pm)|_{x=0},
\eeq
where the integral operators $\hat{v}$ and $\hat{r}^\pm$ possess the kernels
\beq
\label{vt=0}
v(p,q)=\sqrt{\nu(p)}\frac{\sin\frac{l+1}2 (p-q)+\exp[i(\Theta-\frac{p+q}2)]
\sin\frac{l-2}2 (p-q)}{\sin\frac 12 (p-q)}\sqrt{\nu(q)},
\eeq
\beq
\label{rt=0}
r^+(p,q)=\sqrt{\nu(p)}e^{i(\Theta-q)}e^{i\frac l2(p+q)}\sqrt{\nu(q)},
\eeq 
\beq
\label{rt=0-}
r^-(p,q)=\sqrt{\nu(p)}e^{-i\frac l2(p+q)}\sqrt{\nu(q)}.
\eeq 

Comparing the results obtained in this paper with the representations
for the correlation functions of the other models (\cite{l,iik,cikt})
one can see that these representations are rather different and there
are new difficulties in our formulae. The models considered before
were the free fermion points of the more general integrable models.
The phase model being a limiting case of the $q$-boson hopping model
is not a free fermion point of this model and this can explain the
new difficulties appeared in the correlation functions. The very peculiar
property of the correlation functions of the phase model is the dependence
on the full density of particles. 

 This work was supported by the grant number 95-01-00476a of the Russian
Foundation of Fundamental Research. A.I. and N.K. thank Ecole Normale
Superieure de Lyon for hospitality.
N.B. would like to acknowledge the hospitality at the
Research Institute for Theoretical Physics
University of Helsinki.

\end{document}